\documentclass[pdflatex,sn-mathphys-num]{sn-jnl}

\usepackage{graphicx}%
\usepackage{multirow}%
\usepackage{amsmath,amssymb,amsfonts}%
\usepackage{amsthm}%
\usepackage{mathrsfs}%
\usepackage[title]{appendix}%
\usepackage{xcolor}%
\usepackage{textcomp}%
\usepackage{manyfoot}%
\usepackage{booktabs}%
\usepackage{algorithm}%
\usepackage{algorithmicx}%
\usepackage{algpseudocode}%
\usepackage{listings}%

\usepackage{isotope}
\usepackage{graphicx,tabularx,booktabs,colortbl}
\usepackage{pgfplotstable}
\usepackage{subcaption}
\usepackage[T1]{fontenc}
\usepackage{hyperref}

\theoremstyle{thmstyleone}%
\theoremstyle{thmstyletwo}%

\theoremstyle{thmstylethree}%

\newcommand{\eq}{\; = \;}
\newcommand{\ps}{\mbox{ps}}
\newcommand{\ns}{\mbox{ns}}
\newcommand{\mm}{\mbox{mm}}
\newcommand{\kev}{\mbox{keV}}
\newcommand{\mev}{\mbox{MeV}}

\newcommand{\ga}{\isotope[68]{Ga}}
\newcommand{\iod}{\isotope[124]{I}}
\newcommand{\nai}{[\isotope[124]{I}]NaI}
\newcommand{\rb}{\isotope[82]{Rb}}

\newcommand{\tops}{ \tau_{3} }
\newcommand{\tge}{$3\gamma\mbox{E}$}

\begin{document}

\title[Positronium Lifetime Imaging with Biograph Vision Quadra]{Positronium Lifetime Imaging with the Biograph Vision Quadra using \textsuperscript{124}I}


\author*[1,2]{\fnm{Lorenzo} \sur{Mercolli}}\email{lorenzo.mercolli@insel.ch}
\author[3]{\fnm{William M.} \sur{Steinberger}}\email{william.steinberger@siemens-healthineers.com}
\author[1,2]{\fnm{Narendra} \sur{Rathod}}\email{narendra.rathod@unibe.ch}
\author[3]{\fnm{Maurizio} \sur{Conti}}\email{maurizio.conti@siemens-healthineers.com}
\author[4,5]{\fnm{Paweł} \sur{Moskal}}\email{p.moskal@uj.edu.pl}
\author[1]{\fnm{Axel} \sur{Rominger}}\email{axel.rominger@insel.ch}
\author[1]{\fnm{Robert} \sur{Seifert}}\email{robert.seifert@insel.ch}
\author[1,2]{\fnm{Kuangyu} \sur{Shi}}\email{kuangyu.shi@insel.ch}
\author[4,5]{\fnm{Ewa Ł.} \sur{Stępień}}\email{e.stepien@uj.edu.pl}
\author[1,2,6]{\fnm{Hasan} \sur{Sari}}\email{hasan.sari@unibe.ch}

\affil*[1]{\orgdiv{Department of Nuclear Medicine}, \orgname{Inselspital, Bern University Hostpital, University of Bern}, \orgaddress{\city{Bern}, \country{Switzerland}}}

\affil[2]{\orgdiv{ARTORG Center for Biomedical Engineering Research}, \orgname{University of Bern}, \orgaddress{\city{Bern}, \country{Switzerland}}}

\affil[3]{\orgname{Siemens Medical Solutions USA, Inc.}, \orgaddress{\city{Knoxville TN}, \country{USA}}}

\affil[4]{\orgdiv{Faculty of Physics, Astronomy and Applied Computer Science}, \orgname{Jagiellonian University}, \orgaddress{\city{Krakow}, \country{Poland}}}

\affil[5]{\orgdiv{Centre for Theranostics}, \orgname{Jagiellonian University}, \orgaddress{\city{Krakow}, \country{Poland}}}

\affil[6]{\orgname{Siemens Healthineers International AG}, \orgaddress{\city{Z{\"u}rich}, \country{Switzerland}}}



\abstract{\textbf{Purpose:} Measuring the ortho-positronium (oPs) lifetime in human tissue bears the potential of adding clinically relevant information about the tissue microenvironment to conventional positron emission tomography (PET). Through phantom measurements, we investigate the voxel-wise measurement of oPs lifetime using a commercial long-axial field-of-view (LAFOV) PET scanner.   
 
\textbf{Methods:} We prepared four samples with mixtures of Amberlite XAD4, a porous polymeric adsorbent, and water and added between 1.12 MBq and 1.44 MBq of \textsuperscript{124}I. The samples were scanned in two different setups: once with a couple of centimeters between each sample (15 minutes scan time) and once with all samples taped together (40 minutes scan time). For each scan, we determine the oPs lifetime for the full samples and at the voxel level. The voxel sizes under consideration are 10.0\textsuperscript{3} mm\textsuperscript{3}, 7.1\textsuperscript{3} mm\textsuperscript{3} and 4.0\textsuperscript{3} mm\textsuperscript{3}.
 
\textbf{Results:} Amberlite XAD4 allows the preparation of samples with distinct oPs lifetime. Using a Bayesian fitting procedure, the oPs lifetimes in the whole samples are 2.52±0.03 ns, 2.37±0.03 ns, 2.27±0.04 ns and 1.82±0.02 ns, respectively. The voxel-wise oPs lifetime fits showed that even with 4.0\textsuperscript{3} mm\textsuperscript{3} voxels the samples are clearly distinguishable and a central voxels have good count statistics. However, the situation with the samples close together remains challenging with respect to the spatial distinction of regions with different oPs lifetimes.  
 
\textbf{Conclusion:} Our study shows that positronium lifetime imaging on a commercial LAFOV PET/CT should be feasible under clinical conditions using \textsuperscript{124}I. 
}

\keywords{Positronium lifetime imaging, Long axial field-of-view PET/CT, \textsuperscript{124}I}




\maketitle

\section*{Introduction}

Ortho-positronium (oPs), the spin 1 state of an electron-positron bound state, has a significantly longer lifetime in vacuum than the spin 0 state, which is called para-positronium (pPs). The lifetime of pPs is too short in order to interact significantly with the environment~\cite{bass2023}. However, oPs has a lifetime of about $142 \ns$ in vacuum and it can therefore undergo different interactions with surrounding atoms and molecules (see e.g.\ Refs.~\cite{vertes2011,moskal2019b,bass2023,hourlier2024}). In particular, the oPs' positron can annihilate with an environmental electron, and thereby, the oPs lifetime can be significantly shortened. This so-called pick-off process makes the oPs lifetime dependent on the atomic and molecular structure of the surrounding material. oPs lifetime is also shortened by a spin exchange process depending on the concentration of oxygen molecules~\cite{shibuya2020,bass2023,moskal2022}. In vacuum oPs decays into three photons, while in matter due to the pick-off and conversion processes, it may annihilate also into two photons. In principle, both decays can be used for measuring the lifetime of oPs properties. However, it was shown that oPs lifetime imaging based on the two-photon annihilation is 300 times more efficient than oPs imaging based on annihilation into three photons~\cite{moskal2020b,moskal2020b,moskal2022frontiers}.

There is a significant interest in the medical domain for oPs lifetime measurements (see e.g.\ Refs.~\cite{moskal2019,moskal2019b,moskal2020b,bass2023,hourlier2024,tashima2024}), mainly driven by the possibility to measure oxygenation levels in human tissue, \cite{shibuya2020,Stepanov2020,Stepanov2021,moskal2022,takyu2024b,takyu2024}, to asses tissue pathology in vivo ~\cite{moskal2021,karimi2023,moskal2023,moskal2024brain,avachat2024} and to sense pH level and electrolytes within the tissue~\cite{shimazoe2022,shimazoe2022b,zaleski2023,takyu2024b,shimazoe2024}. Recently, the first in vivo positronium images~\cite{moskal2024brain} and the first in-vivo measurements of oPs lifetime with clinical positron emission tomography (PET) system~\cite{moskal2024brain,mercolli2024} were demonstrated. The oPs lifetime has the potential to add diagnostic information, which is currently unavailable or requires additional interventions, such as e.g.\ biopsy or additional use of hypoxia tracers.   

In Refs.~\cite{steinberger2024,mercolli2024}, we showed that it is possible to do oPs lifetime measurements with a commercial long axial field-of-view (LAFOV) PET scanner \cite{Prenosil2022,Spencer2021}. However, in Ref.~\cite{mercolli2024} we also showed that the collection of sufficient count statistics with a PET scanner is a major challenge. The voxel-wise determination of oPs lifetime, what is usually called oPs lifetime imaging, has been shown to be feasible only with long-lived radionuclides, long scan times, large voxel sizes or simplifying the fit models \cite{moskal2024brain,moskal2021,chen2024,moskal2020,moskal2019ieee,shopa2023}. Usually, a combination of multiple of these methods is required.   

In this report, we show that oPs lifetime imaging can be achieved using a commercial PET/CT scanner under conditions typically encountered in clinical practice with respect to isotope, activity concentration, scan time, and voxel size.
As highlighted in Refs.~\cite{steinberger2024,mercolli2024,takyu2023}, \iod\ possesses favorable characteristics for oPs lifetime imaging~\cite{takyu2023}, it is also well suited for oPs imaging with the Biograph Vision Quadra (Siemens Healthineers, USA)~\cite{steinberger2024,mercolli2024} and is routinely used in some departments due to the favorable imaging characteristics
compared to conventional \isotope[131]{I} imaging. In contrast to other thyroid-directed PET tracers, like [\isotope[18]{F}]tetrafluoroborate, \iod\ PET has usually higher uptake and also enables delayed imaging which can be used for dosimetry applications~\cite{plyku2022,dittmann2020,ventura2023}. Using phantom measurements, we identify the conditions under which oPs lifetime imaging is viable and bring to the fore the remaining challenges.

\section{Materials and methods}

\begin{figure}
    \centering
    \includegraphics[width=0.8\linewidth]{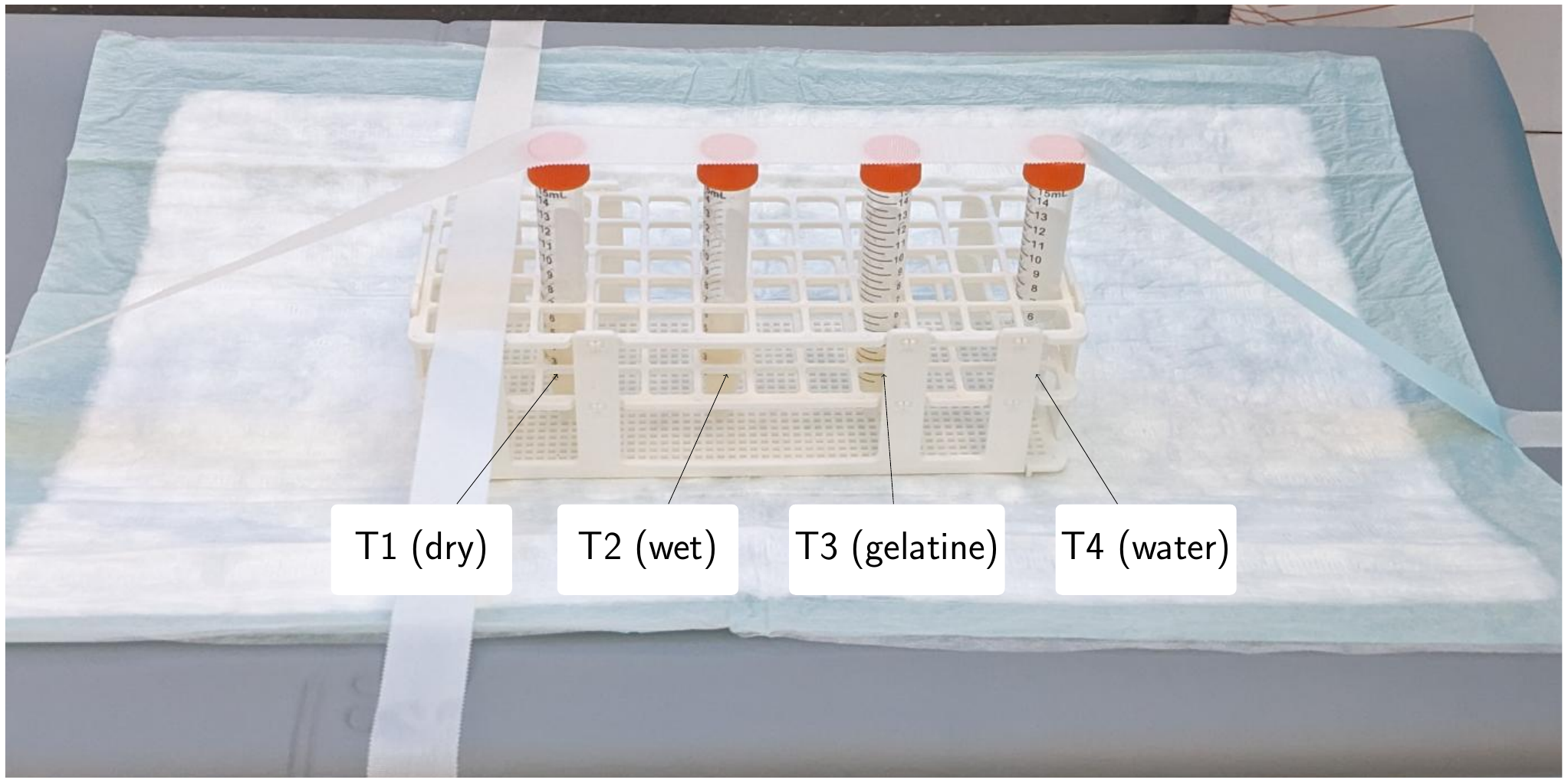}
    \caption{Picture of the experimental setup with the four samples separated from each other}
    \label{f:experimental_setup}
\end{figure}

In order to asses the capabilities of Quadra with respect to oPs lifetime imaging, we filled four chemistry tubes with different mixtures of Amberlite XAD4 (Sigma-Aldrich, Co., St. Louis MO, USA) and demineralized water. As shown in Ref.~\cite{lapkiewicz2022}, XAD4 allows to vary the oPs lifetime with a simple experimental setup. A relatively low activity of \nai\ was added to each tube. Tab.~\ref{t:samples} summarizes the details of the sample preparation and Fig.~\ref{f:experimental_setup} shows the experimental setup. The first tube contained XAD4 that was air-dried for 24 hours. T2 cointained the wet XAD4 (as it is delivered), while for T2 we added $1 \, \mbox{ml}$ of gelatine to $3.5 \, \mbox{ml}$ wet XAD4. In Tab.~\ref{t:samples} $m_\text{XAD4}$ is the weight of the wet XAD4 and the gelatine together. To all tubes, about $0.4\, \mbox{ml}$ \nai\ solution was added.


\begin{table}
    \centering
    \caption{Summary of the sample preparation. $0.4\, \mbox{ml}$ \nai\ was added to all tubes. $\rho_\text{tot}$ contains the \nai. }
    \label{t:samples}
    \begin{tabular}{llcccc}
        \toprule
        Sample & XAD4 & $A \; [\mbox{MBq}]$ &  $V_\text{XAD4} \, [\mbox{ml}]$ &  $m_\text{XAD4} \, [\mbox{g}]$ & $\rho_\textbf{tot} \, [\mbox{g}/\mbox{cm}^3]$  \\
        \midrule
        T1 & Dry & 1.12 &  5.0 & 3.25 & 0.67 \\
        \rowcolor{lightgray}T2 & Wet & 1.44  &  5.5 & 3.56 & 0.67 \\
        T3 & Gelatine & 1.14 &  4.5 & 3.49 & 0.79 \\
        \rowcolor{lightgray}T4 & Deminalized water & 1.26 &  5.0 & 5.0 & 1.0 \\
        \bottomrule
    \end{tabular}
\end{table}

\begin{figure}[htb]
    \centering
    \includegraphics[width=0.8\linewidth]{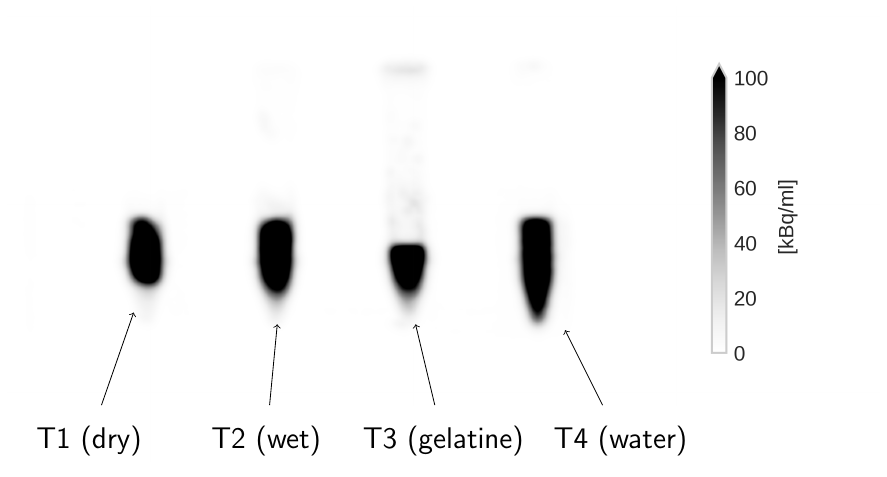}
    \caption{MIP of the coincidence PET images of the four separated tubes. }
    \label{f:pet}
\end{figure}

\begin{figure}
    \centering
    \includegraphics[width=0.49\linewidth]{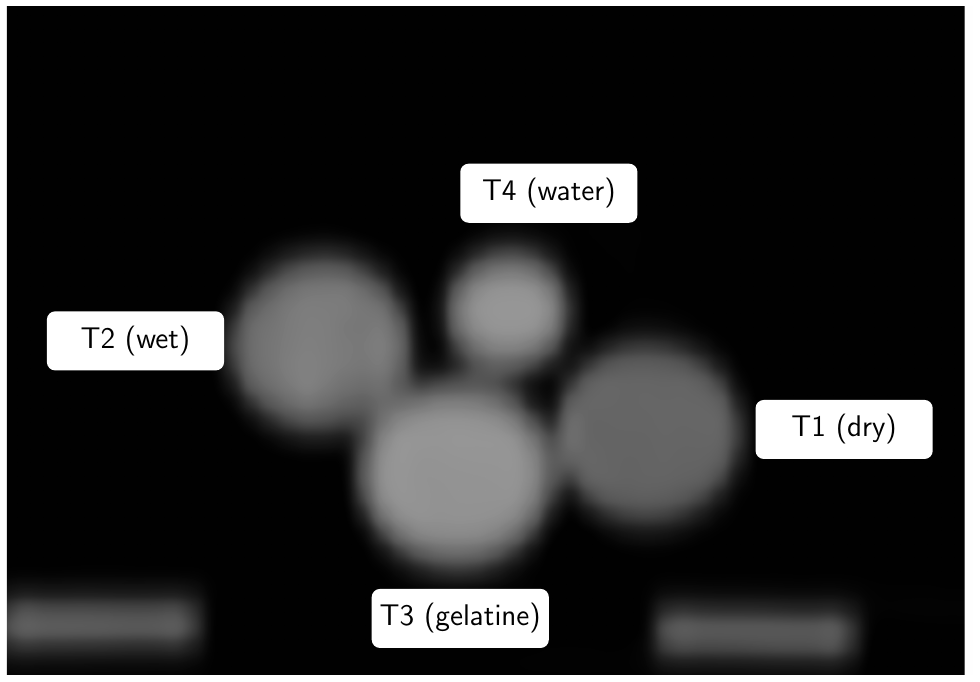}
    \includegraphics[width=0.49\linewidth]{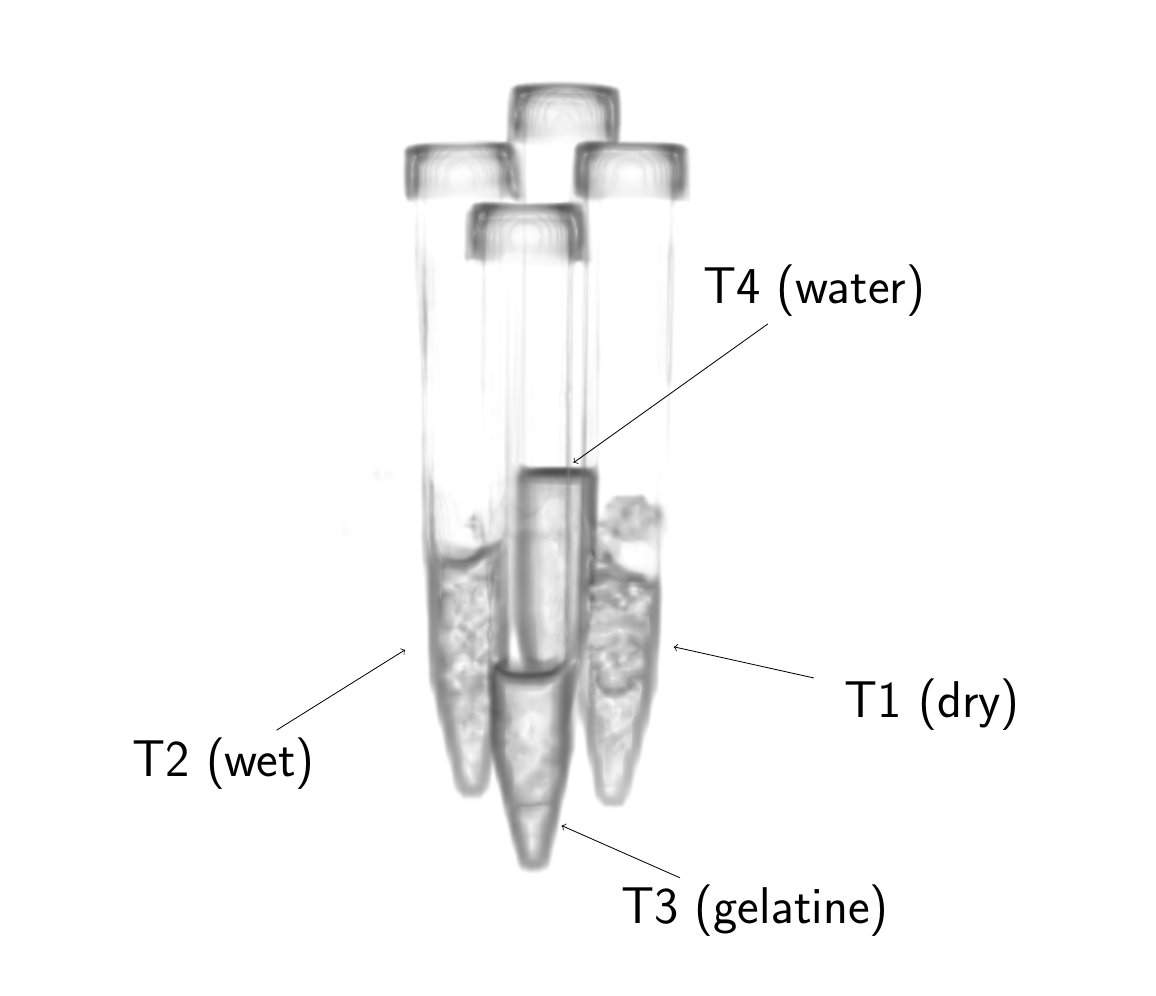}
    \caption{Top view of a CT slice (right) and 3D rendering of the CT (left) of the setup with the four tubes taped together. }
    \label{f:ct_close}
\end{figure}

The samples were measured once with a large distance between them, as shown in Fig.~\ref{f:experimental_setup}, and once taped together. In Fig.~\ref{f:pet} we show the maximum intensity projection (MIP) of the coincidence PET image of the setup with separated tubes (5 minutes scan time). The voxel size is $1.65\times1.65\times1.65 \, \mm^3$.
The experimental setup with the tubes taped together is depicted Fig.~\ref{f:ct_close} with the top view of a CT slice and a 3D rendering from the CT. In the CT images, the voxel size is $1.52\times1.52\times1.65 \, \mm^3$.

In addition, we measured the samples 15 minutes (separated) and 40 minutes (taped together) in singles mode. As described in Ref.~\cite{steinberger2024}, singles mode in Quadra records all single-crystal interactions into a list mode file. A prototype software was used to sort all three-photon events (\tge), i.e.\ events with two photons in the annihilation window $[460\, \kev,545\, \kev] $ and one photon in the prompt energy window of $[568\, \kev,639\, \kev] $. \iod\ has the convenient property of having a prompt photon with an energy of $602.73 \pm 0.08 \, \kev$, which Quadra's detector can fully resolve, and a relatively high branching ratio of $62.9 \pm 0.7 \,\%$ \cite{nds124,das2023}. 
The spatial location of a \tge\ is determined through the time-of-flight (TOF) information of the two annihilation photons, i.e.\ there is no image reconstruction along the line of Refs.~\cite{qi2022,Shibuya2022,chen2023,Huang2024}. The histoimages of \tge\ are binned into three different voxel sizes of $10.0\times 10.0 \times 10.0 \, \mm^3$, $7.1\times 7.1 \times 7.1 \, \mm^3$ and $4.0\times 4.0 \times 4.0 \, \mm^3$, respectively. We consider the $10\times10\times10 \, \mbox{mm}^3$ voxel size to be on the verge of clinical usefulness. It is also the same order of magnitude as the maximum positron range in water for \iod\ (continuous slowing down approximation range for a $2\,\mev$ positron in water is $9.8\,\mbox{mm}$ according to NIST PSTAR). On the other end, we chose the smallest voxel size such that it would really push oPs imaging on Quadra to its limits. According to Refs.~\cite{kertesz2022,kersting2023}, $4\,\mbox{mm}$ is about the spatial resolution that is achievable in coincidence PET imaging with \iod. Finally, $7.1\times7.1\times7.1 \, \mbox{mm}^3$ sits in the middle and could be thought of as similar to the spatial resolution of a SPECT/CT system. As a comparison, we also perform a fit that encompasses all \tge\ in a single tube.  

The time difference distributions (TDD) are the binned time differences between the annihilation and prompt photons in each voxel (time bin width is $133\, \mbox{ps}$, i.e. slightly above Quadras time resolution). We use the Bayesian fitting procedure discussed in Refs.~\cite{steinberger2024,mercolli2024} to determine the oPs lifetime $\tops$ from a measured TDD. The fitting model is the same as in Refs.~\cite{steinberger2024,mercolli2024}, i.e.\ a Gaussian function convoluted with three lifetime components for pPs, direct annihilation and oPs. In contrast to Ref.~\cite{mercolli2024}, we fix the pPs lifetime $\tau_1 = 125 \,\ps$ and direct annihilation $\tau_2= 388\, \ps$ together with the background count number. The background is fixed as the mean value of time differences that are smaller than $-2.5 \, \ns$. We fit the following priors with a Gaussian likelihood to the voxel-wise TDD

\begin{equation}
    \begin{aligned}
        & \tops \sim \mathcal{N}(1.78 \, \ns, 0.8\, \ns) \;, \\
        & BR_{1,2,3} \; \sim \; \mathrm{Dirichlet}(0.75, 3.1, 1.15) \;,\\
        & \sigma   \sim  \mathcal{N}(0.1 \,\ns, 0.05  \,\ns) \,, \\
        & \Delta  \sim \mathcal{N}(0 \,\ns, 0.5 \,\ns) \;, \\
        & N \; \sim \; \mathcal{N}(A, 0.1\cdot A) \quad \mbox{with} \quad A \eq \int dt \, \left( y_i - b \right) \;, 
    \end{aligned}
\end{equation}
where $b$ is the background value and $y_i$ are the bin values of the TDD. In the oPs lifetime images, we only selected voxels that have a relative error in the background region of less than $20\%$. We fit time differences in the range from $-2\,\ns$ to $8\,\ns$.

The posterior distribution of $\tops$ is bell-shaped with hardly any skewness. We therefore report the uncertainty of $\tops$ with a standard deviation, which is estimated with the common point estimate. The uncertainty of the branching ratios $BR_{1,2,3}$ is given in terms of the $68\%$ highest density intervals (HDI) of the posterior distribution since a point estimate of the standard deviation would not make sense for a Dirichlet variable.

\section{Results}

\begin{figure}
    \centering
    \includegraphics[width=0.9\linewidth]{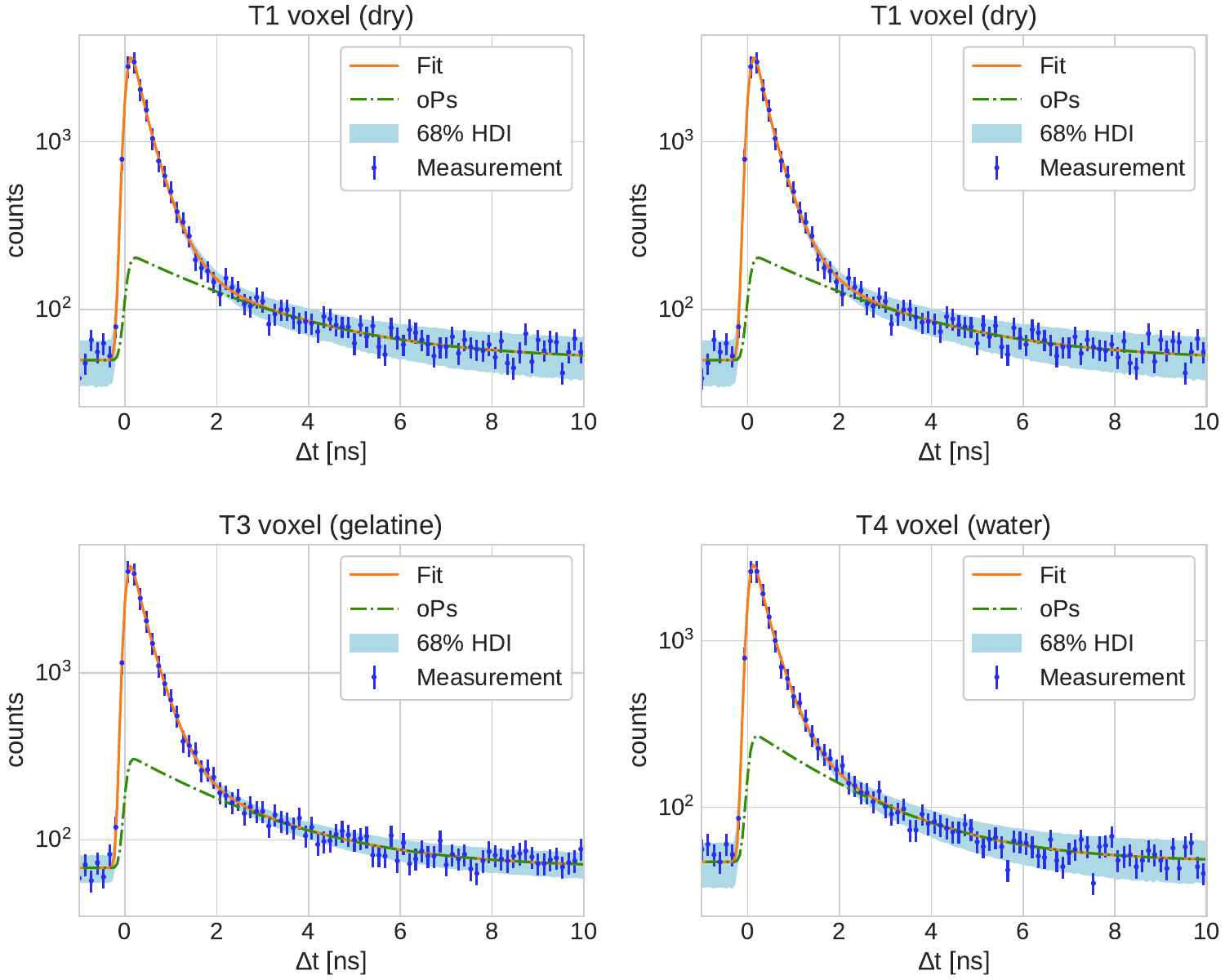}
    \caption{Single-voxel TDD together with the fit prediction, posterior 68\% HDI and oPs component in logarithmic scale ($4\times4\times4 \, \mm^3$ voxel size). The fit results are reported in Tab.~\ref{t:fit_tubes}. }
    \label{f:single_tdd}
\end{figure}

The top four rows of Tab.~\ref{t:fit_tubes} report the single tube fits, i.e.\ when all measured time differences in a tube are collected in one TDD (no spatial binning of the \tge\ data). Clearly, the different humidity levels of the XAD4 powder lead to significantly distinct oPs lifetimes. Furthermore, Tab.~\ref{t:fit_tubes} includes also the results from fitting a TDD from a single $4\times4\times4 \, \mm^3$ voxel. The voxel is chosen in the central region of each tube. This allows us to get a good intuition about the count statistics for the smallest voxel size. Fig.~\ref{f:single_tdd} shows the corresponding single-voxel TDD together with the fit prediction for the oPs lifetime component. 

\begin{table}[]
    \centering
    \caption{Fit results for the full samples (top four rows) and single voxel with a size of $4\times4\times4 \, \mm^3$ (bottom four rows). } \label{t:fit_tubes}
    \pgfplotstabletypeset[col sep=semicolon,
        columns/Sample/.style={column type=l,column name=Sample, string type},
        columns/τoPs/.style={column type=l,column name=$\tops \, [\ns]$, string type},
        columns/HDIτ/.style={column type=l,column name=$\mbox{HDI}_{\tops} \, [\ns]$, string type},
        columns/BR1/.style={column type=c,column name=$BR_1$, string type},
        columns/HDIBR1/.style={column type=c,column name=$\mbox{HDI}_{BR_1}$, string type},
        columns/BR2/.style={column type=c,column name=$BR_2$, string type},
        columns/HDIBR2/.style={column type=c,column name=$\mbox{HDI}_{BR_2}$, string type},
        columns/BR3/.style={column type=c,column name=$BR_3$, string type},
        columns/HDIBR3/.style={column type=c,column name=$\mbox{HDI}_{BR_3}$, string type},
        every head row/.style={before row=\toprule,after row=\midrule},
        every last row/.style={after row=\bottomrule},
        every odd row/.style={before row=\rowcolor{lightgray}},
    ]{Sample;τoPs;BR1;HDIBR1;BR2;HDIBR2;BR3;HDIBR3
T1 (dry);2.52±0.03;0.073;[0.071, 0.074];0.716;[0.714, 0.718];0.211;[0.21, 0.213]
T2 (wet);2.37±0.03;0.081;[0.079, 0.082];0.702;[0.7, 0.704];0.217;[0.216, 0.219]
T3 (gelatine);2.27±0.04;0.073;[0.071, 0.075];0.705;[0.702, 0.708];0.222;[0.22, 0.223]
T4 (water);1.82±0.02;0.088;[0.087, 0.09];0.644;[0.641, 0.646];0.268;[0.267, 0.269]
T1 voxel;2.56±0.23;0.09;[0.081, 0.099];0.702;[0.687, 0.716];0.208;[0.2, 0.216]
T2 voxel;2.37±0.23;0.073;[0.062, 0.085];0.69;[0.671, 0.708];0.237;[0.227, 0.247]
T3 voxel;2.3±0.12;0.044;[0.039, 0.05];0.745;[0.736, 0.754];0.21;[0.205, 0.215]
T4 voxel;2.0±0.14;0.062;[0.052, 0.072];0.683;[0.666, 0.7];0.255;[0.246, 0.264]
}
\end{table}

\begin{figure}
    \centering
    \includegraphics[width=\linewidth]{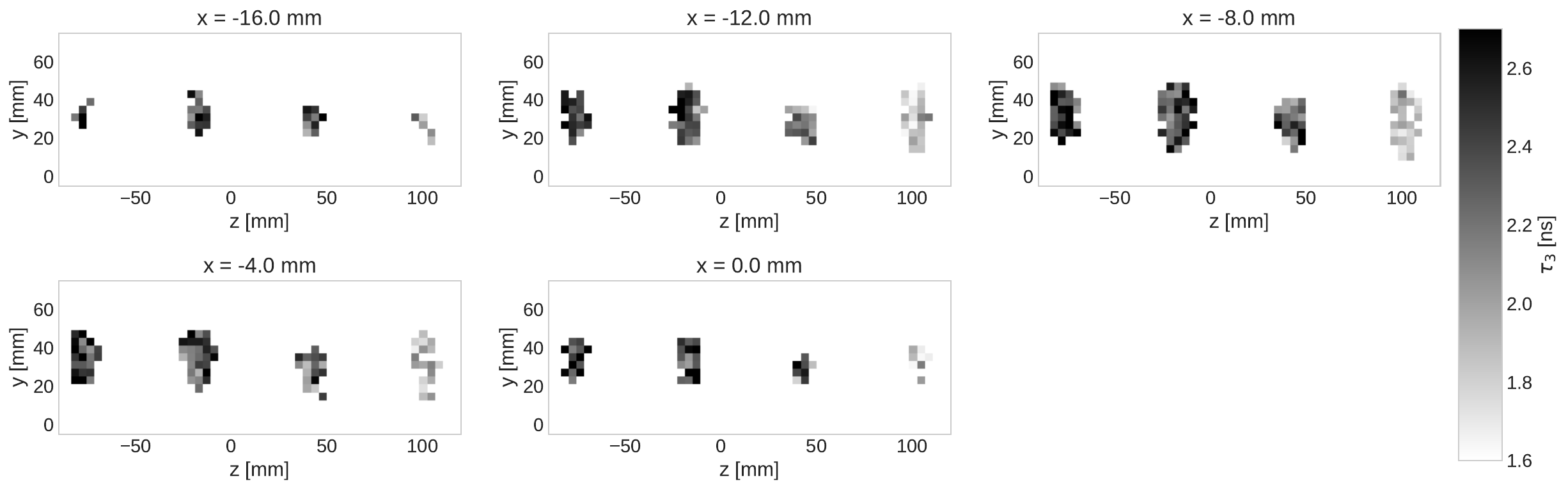}\\
    \vspace{4ex}
    \includegraphics[width=\linewidth]{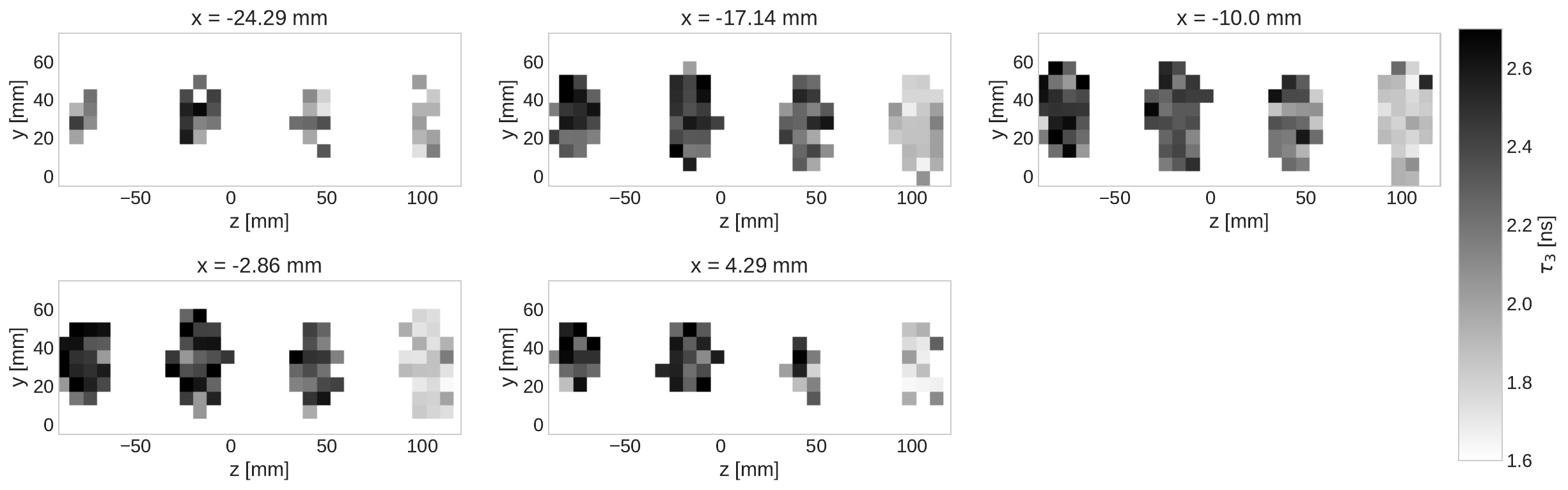}\\
    \vspace{4ex}
    \includegraphics[width=\linewidth]{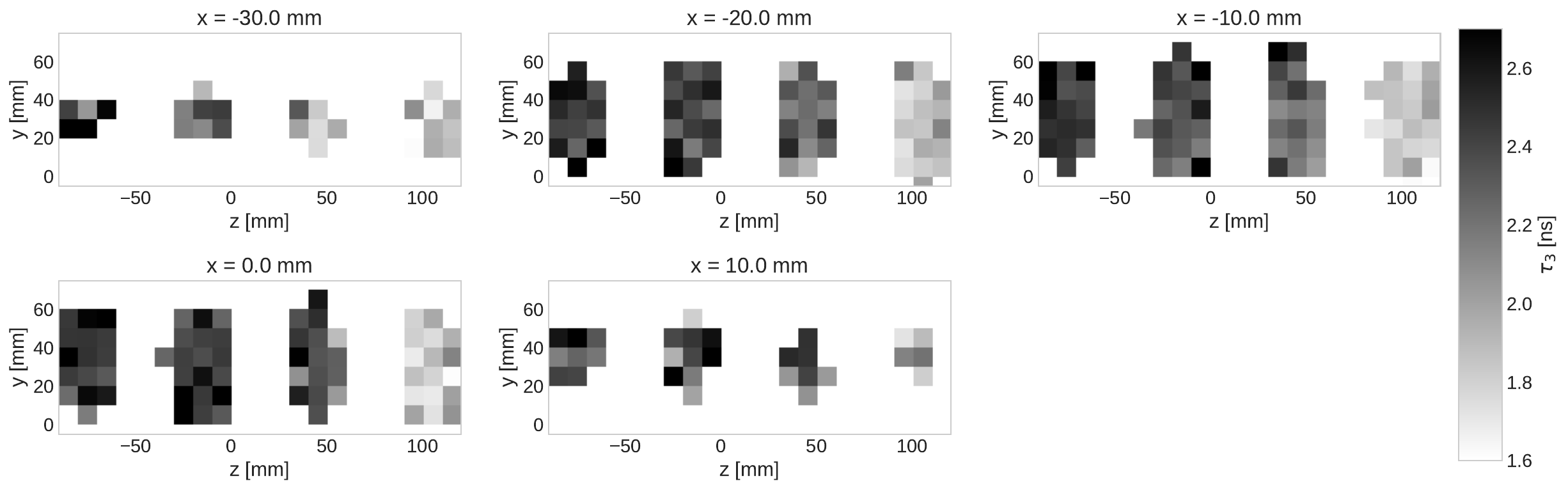}
    \caption{Slices of the oPs lifetime images along the $x$-axis for the separated tubes. The voxel size is increasing from top to bottom. }
    \label{f:slices_sep}
\end{figure}

\begin{figure}
    \centering
    \includegraphics[width=0.85\linewidth]{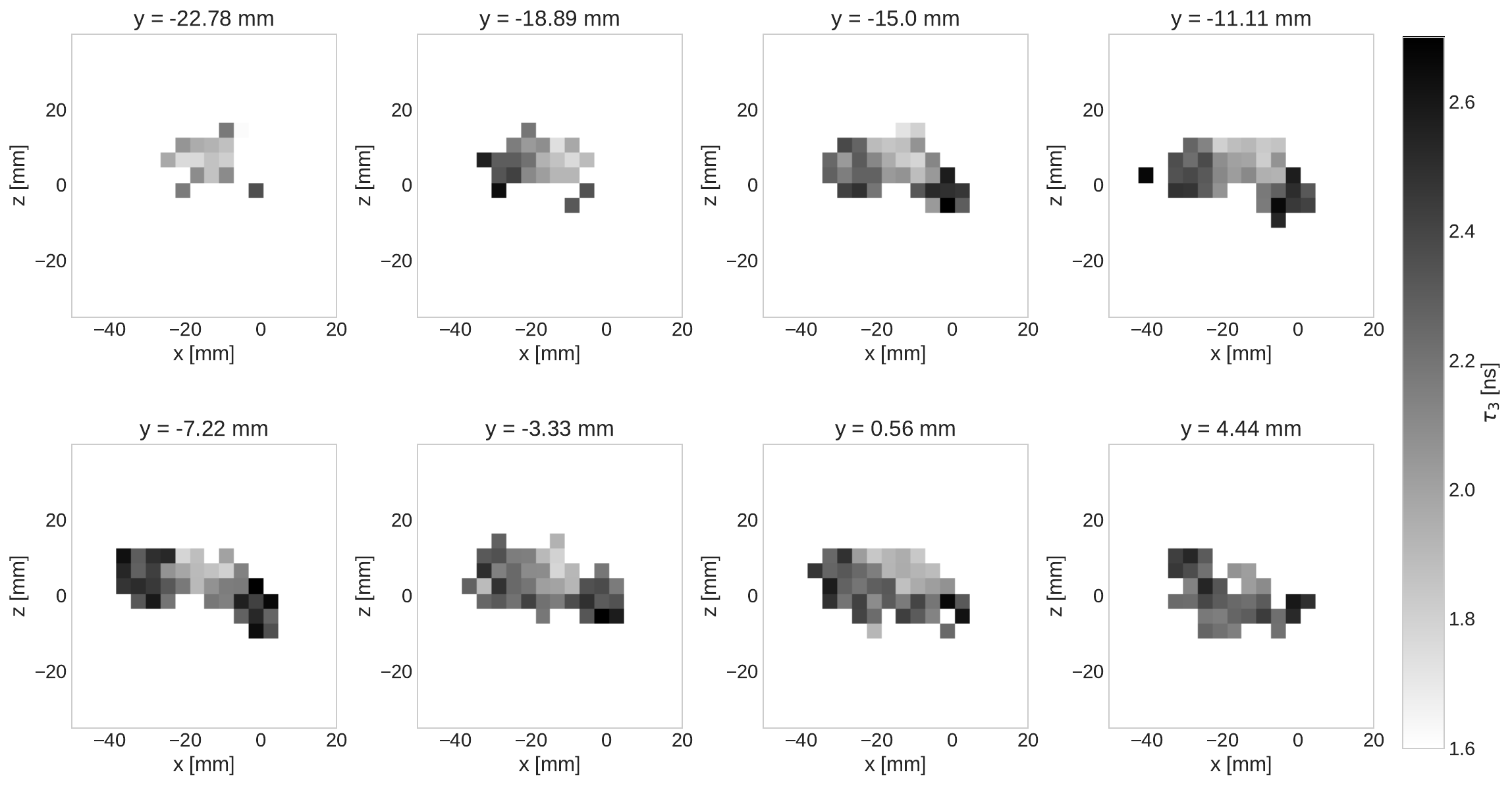}\\
    \vspace{4ex}
    \includegraphics[width=0.85\linewidth]{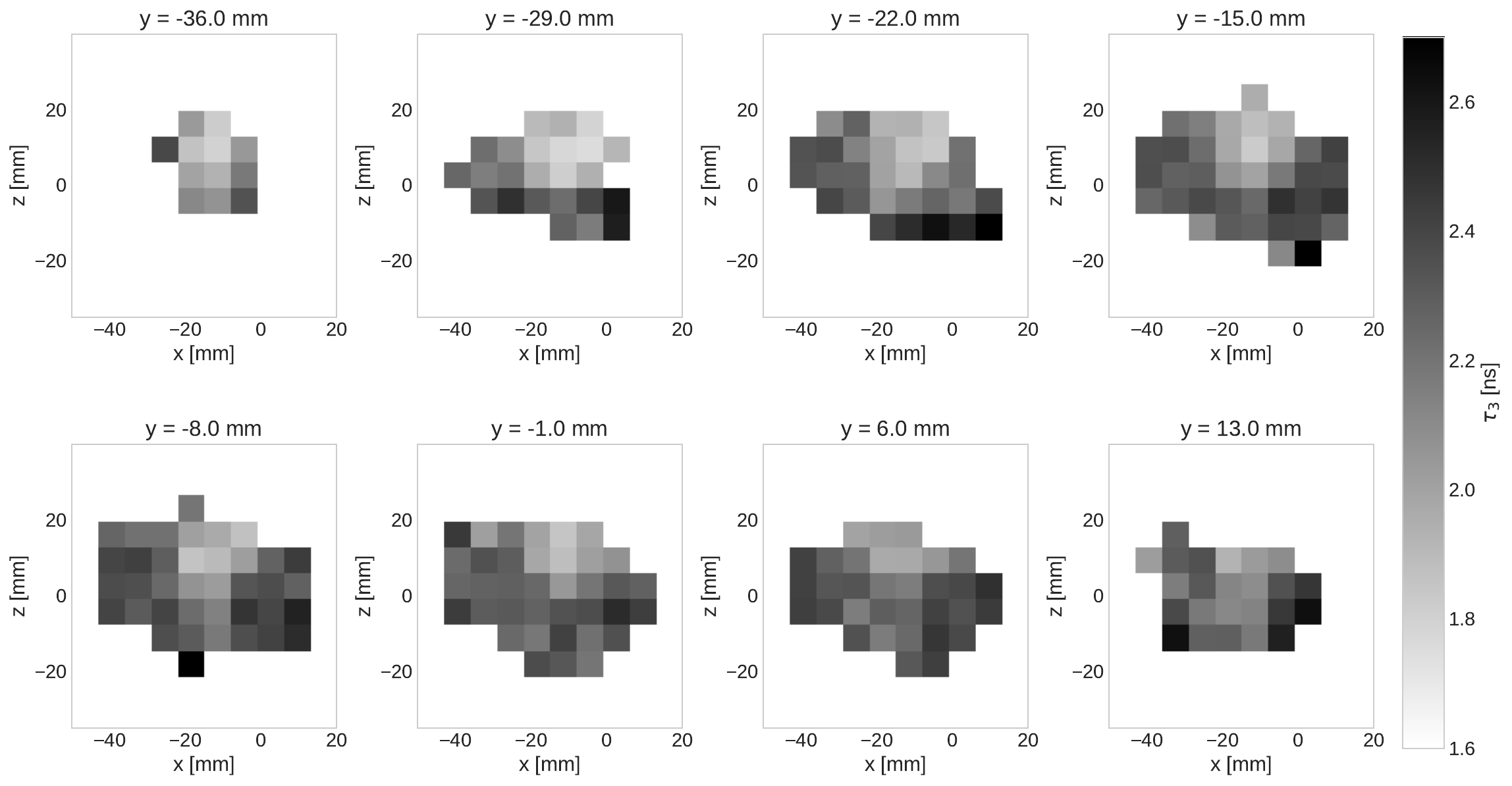}\\
    \vspace{4ex}
    \includegraphics[width=0.85\linewidth]{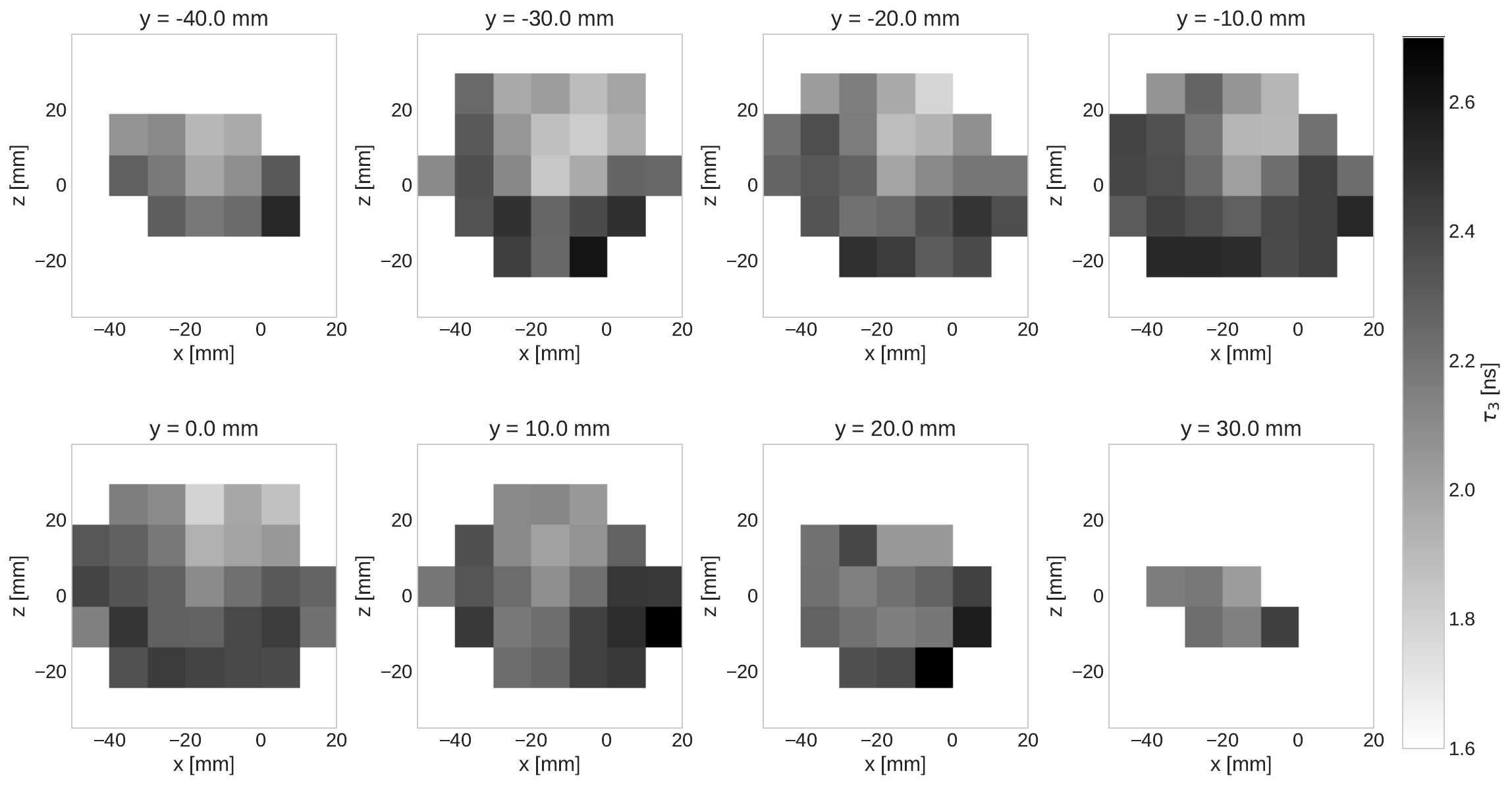}
    \caption{Slices of the oPs lifetime images along the $y$-axis for the tubes taped together. The voxel size is increasing from top to bottom. }
    \label{f:slices_close}
\end{figure}

In Figs.~\ref{f:slices_sep} and \ref{f:slices_close} we show 2D slices of the oPs lifetime images for the two scans with separated and taped-together tubes. The two top rows show slices for the $4\times4\times4 \, \mm^3$ voxel size, while the middle and bottom rows are for the $7.1\times7.1\times7.1 \, \mm^3$ and $10\times10\times10 \, \mm^3$ voxel sizes, respectively. For best visualization, the slices of the separated tubes in Fig.~\ref{f:slices_sep} are shown in the $z-y$ plane, analogously to a coronal PET MIP in Fig.~\ref{f:pet}. For the tubes close together, we chose the $x-y$ plane as in the CT slice in Fig.~\ref{f:ct_close}. No post-processing, such as smoothing or any filtering, was applied to the oPs lifetime images.

\begin{figure}
    \centering
    \includegraphics[width=\linewidth]{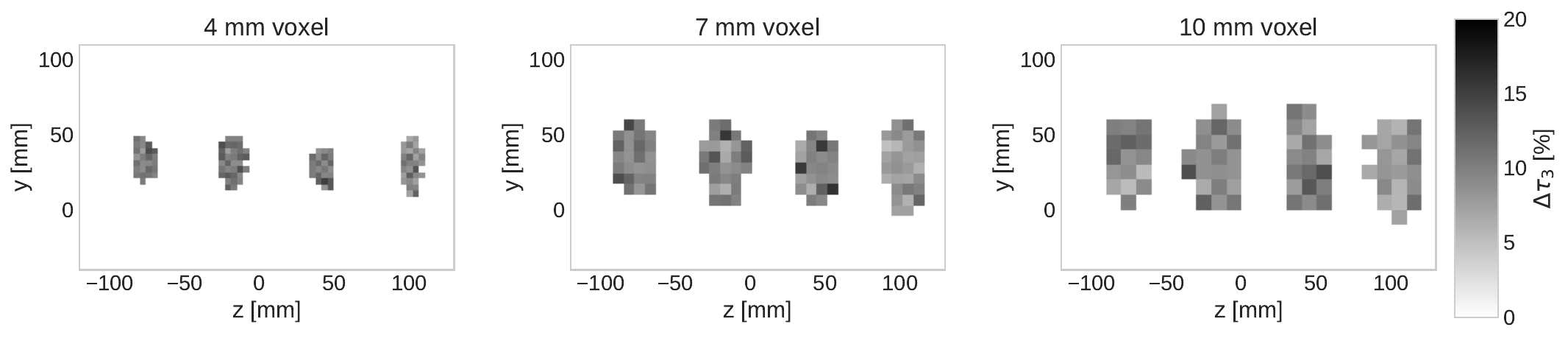} \\
    \vspace{4ex}
    \includegraphics[width=0.8\linewidth]{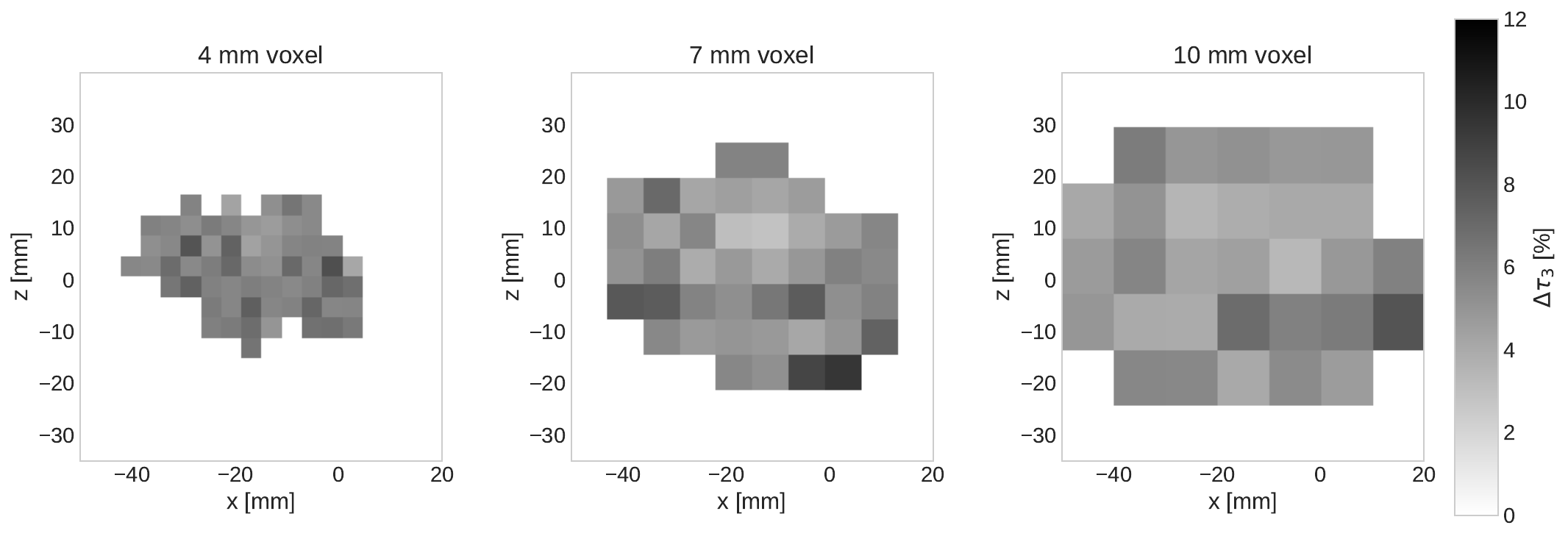}
    \caption{MIP of the relative uncertainties of $\tops$ for the two separated (top) and taped together (bottom) experimental setups.}
    \label{f:error_mip}
\end{figure}

Finally, we show the MIP of the standard deviation for $\tops$ for the three voxel sizes and the two experimental setups in Fig.~\ref{f:error_mip}. The color bar in these figures shows the relative uncertainty of $\tops$ for each voxel. 

\section{Discussion}

First, we would like to highlight the low statistical uncertainty of the oPs lifetime in the single tube fit in Tab.~\ref{t:fit_tubes}. Within a few ml, an activity concentration low as $232 \, \mbox{kBq}/\mbox{ml}$ and a scan time of 15 minutes, the marginalized posterior distribution of $\tops$ has a relative standard deviation of less than $1.76\%$ using a commercial LAFOV PET/CT and Ref.~\cite{steinberger2024}'s methodology. Apart from the higher prompt photon branching ratio of \iod, the main improvement compared \rb\ and \ga\ (see Refs.~\cite{steinberger2024,mercolli2024}) is Quadra's capability to resolve \iod's photopeak and thereby increasing the peak signal-to-background ratio (pSBR). In addition, the smaller decay constant \iod\ allows for a longer scan time and larger time-integrated counts. This high statistical precision allows to distinguish the oPs lifetime in the four samples, confirming Ref.~\cite{lapkiewicz2022} in that XAD4 is well suited for performance evaluations and intercomparisons of PET/CT scanners for oPs lifetime imaging. 

In the literature, oPs lifetime imaging has shown to be feasible only with \isotope[22]{Na} and very long scan times and/or large voxel sizes. E.g.\ Ref.~\cite{moskal2021} applied a spatial binning of $2\times2\times2 \, \mbox{cm}$ to their tissue sample data. Likewise, Ref.~\cite{moskal2024brain} seems to have a voxel size of multiple cm. It remains unclear, how much viable clinical information such voxel sizes may contain. With \rb\ and \ga, i.e.\ radionuclides that are used in clinical routine, the count statistics and in particular the pSBR are not sufficient for oPs lifetime imaging, given our methodology. This is why in Refs.~\cite{steinberger2024,mercolli2024} we refrained from performing voxel-wise oPs lifetime fits.   

Considering Fig.~\ref{f:slices_sep}, it is clear that even with the $4\times4\times4 \, \mbox{mm}^3$ the oPs lifetime in the four tubes are distinguishable. From left to right, $\tops$ decreases according to the sample filling. Homogeneity of $\tops$, i.e.\ of the gray scale, within a single tube clearly gets worse with decreasing voxel size, which simply reflects the decreasing count statistics within a voxel (see also Fig.~\ref{f:error_mip}). Likely, the localization of \tge\ with TOF introduces some uncertainties and we might see events that stem from the tube walls and possibly even the air. The error of $\tops$ increases towards the tube walls, but as Fig.~\ref{f:error_mip} shows, the relative error still remains around $10\%$. Relaxing the $20\%$ background error condition for fitting voxels could possibly provide a shape of the oPs lifetime images that is more consistent with the real shape of the tubes. 

With the second experimental setup, i.e.\ with the tubes tied together, we wanted to create a very challenging situation for oPs lifetime imaging. Indeed, the top view slices in Fig.~\ref{f:slices_close} do not allow for a clear spatial distinction of the four tubes. With respect to the oPs lifetimes, the shorter $\tops$ of the water tube can be best distinguished from the other tubes in the $7.1\times7.1\, \mbox{mm}^2$ slices. Also in the smallest voxel size, the light gray area is distinguishable, but in our opinion the limit of oPs lifetime imaging with Quadra is reached at this point. The longer scan of this setup (40 min vs.\ 15 min) is noticeable in the lower $\tops$ error, as shown in Fig.~\ref{f:error_mip}. However, the limiting factor for this experiment is likely the localization of the \tge\ with TOF. It should be noted that also in the coincidence PET image (reconstructed with OSEM, TOF, $440\times 440$ matrix size, 4 iterations and 5 subsets, $2\,\mm$ Gauss filter) the tubes are not very well spatially distinguishable. The tube shape is better visible than in Fig.~\ref{f:slices_close}, but partial volume effects blur the small gaps between the tubes. The recently proposed reconstruction methods from Refs.~\cite{qi2022,Huang2024} suggest that voxel sizes smaller than $4\,\mbox{mm}$ should be feasible. It would be interesting to see whether the algorithm of Ref.~\cite{Huang2024} could lead to a substantial improvement on Fig.~\ref{f:slices_close}. 

In view of clinical applications, a smoothing (or more sophisticated post-processing) of the oPs lifetime images would certainly improve the diagnostic value of the image. It should be mentioned, that the activity concentration in the four samples (see Tab.~\ref{t:samples}) is still somewhat higher than one could expect in a thyroid cancer patient. E.g.\ Ref.~\cite{jentzen2008} reports an activity concentration between $10$ and $70 \, \mbox{kBq}/\mbox{ml}$ in differentiated thyroid cancer metastases, which could be a prime target for oPs lifetime imaging due to hypoxia dependence of the tumor differentiation \cite{ma2023}.

\section{Conclusions}


This brief report demonstrates that oPs lifetime imaging, achieved as a 3D image with $\tops$ as voxel values, is feasible using a commercial PET/CT scanner under clinically viable conditions with respect to the isotope, activity concentration, scan time and voxel sizes. The Quadra scanner, in combination with our data analysis methodology, is able to capture oPs lifetimes with notable precision, even at voxel sizes as small as $4.0^3\, \mm^3$. These results affirm that the Quadra can yield distinct, voxel-wise lifetime measurements across various sample compositions, enabling diagnostic-level imaging using \iod-based compounds. 
Future work could focus on advanced reconstruction algorithms and smoothing techniques, potentially enhancing both the diagnostic utility and spatial resolution of oPs lifetime images, especially in challenging setups with closely positioned samples.


\backmatter



\section*{Declarations}

\subsection*{Funding}
This research is partially supported by the grant no. 216944 under the Weave/Lead Agency program of the Swiss National Science Foundation and the National Science Centre of Poland through grant OPUS24+LAP No. 2022/47/I/NZ7/03112 and 2021/42/A/ST2/00423.

\subsection*{Competing interests}

WMS and MC are full-time employees of Siemens Medical Solutions USA, Inc. HS is a part-time employee of Siemens Healthineers International AG. \\
PM is an inventor on a patent related to this work. Patent nos.: (Poland) PL 227658, (Europe) EP 3039453, and (United States) US 9,851,456], filed (Poland) 30 August 2013, (Europe) 29 August 2014, and (United States) 29 August 2014; published (Poland) 23 January 2018, (Europe) 29 April 2020, and (United States) 26 December 2017. AR has received research support and speaker honoraria from Siemens. KS received research grants from Novartis and Siemens and conference sponsorships from United Imaging, Siemens, and Subtle Medical not related to the submitted work. All other authors have no conflict of interests to report. RS has received research/travel support from Boehringer Ingelheim Fund and Else Kr{\"o}ner-Fresenius-Stiftung, as well as travel support and lecture fees from Novartis and Boston Scientific, outside the submitted work.

\subsection*{Ethics approval}

Not applicable.

\subsection*{Data availability}

Evaluated data are available in the Zenodo repository \url{https://doi.org/10.5281/zenodo.13443797}.

\subsection*{Consent to participate}

Not applicable.








\bibliography{overleaf_positronium_bibliography}

\end{document}